\documentclass[aps,prl,reprint]{revtex4-1}
\usepackage[latin9]{inputenc}
\usepackage{color}
\usepackage{amssymb}
\usepackage{graphicx}
\usepackage[english]{babel}
\usepackage[T1]{fontenc}
\usepackage{amsmath}
\usepackage{stmaryrd}
\usepackage{graphicx}

\begin{document}

\title{Guiding Dirac fermions in graphene with a carbon nanotube}

\author{Austin Cheng$^{1}$, Takashi Taniguchi$^{2}$, Kenji Watanabe$^{2}$,
Philip Kim$^{1}$, Jean-Damien Pillet$^{3}$}
\email{jean-damien.pillet@polytechnique.edu}

\affiliation{$^{1}$Department of Applied Physics, Harvard University, Cambridge, MA, USA,}

\affiliation{$^{2}$National Institute for Material Science, Tsukuba, Japan,}

\affiliation{$^{3}$LSI, CEA/DRF/IRAMIS, Ecole Polytechnique, CNRS, Institut Polytechnique
de Paris, F-91128 Palaiseau, France.}

\begin{abstract}
Relativistic massless charged particles in a two-dimensional
conductor can be guided by a one-dimensional electrostatic potential,
in an analogous manner to light guided by an optical fiber. We use
a carbon nanotube to generate such a guiding potential in graphene
and create a single mode electronic waveguide. The nanotube and graphene
are separated by a few nanometers and can be controlled and measured
independently. As we charge the nanotube, we observe the formation
of a single guided mode in graphene that we detect using the same
nanotube as a probe. This single electronic guided mode in graphene
is sufficiently isolated from other electronic states of linear Dirac
spectrum continuum, allowing the transmission of information with
minimal distortion.
\end{abstract}

\maketitle

Like a photon, an electron can be used as a carrier of information \cite{bauerle_coherent_2018}.
However, there is a limited number of tools to control a single electron \cite{zhao_creating_2015}
and the simple fact of guiding it coherently in a solid, like an optical fiber
for light, is a technological feat \cite{williams_gate-controlled_2011,rickhaus_guiding_2015}.
One-dimensional materials such as semiconducting nanowires naturally
provide guidance for electrons, but in these materials, electrons
can only be transmitted over short distances before losing its information
\cite{chuang_ballistic_2013}. Another possibility is through the
edge channel of a two-dimensional electron gas in the quantum Hall
regime, but a large magnetic field is required for the channel to
be a single mode \cite{rickhaus_snake_2015}, which is crucial for
the carried information not to be distorted during propagation.

\begin{figure}
\includegraphics[width=1\columnwidth]{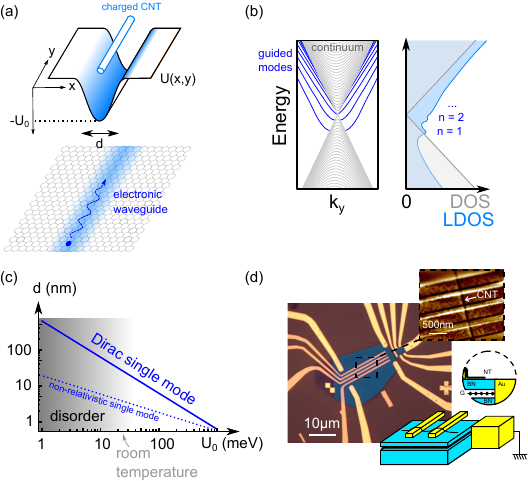}

\caption{\label{fig:Fig1}Electron waveguide in graphene. (a)
Schematic of graphene with a potential well represented by the blue
region which confines electrons along the y-direction. Below, electrostatic
potential along the x-direction generated by a charged carbon nanotube
(CNT). (b) Schematic of the band structure as a function of
momentum $k_{y}$. The grey lines correspond to the bulk states and
the blue lines correspond to the guided modes. On the right: global
density of states (DOS) and local density of states (LDOS) as a function
of energy. (c) Diagram showing the condition (blue line) for
a waveguide to host a Dirac single mode and non-relativistic single
mode. (d) Optical image of one of the devices. EFM picture
of a CNT on top of a h-BN encapsulated graphene device with metallic
electrodes, and a schematic of the device structure.}
\end{figure}

An alternative approach, conceptually similar to an optical fiber \cite{gloge_weakly_1971},
is to use an electrostatic potential well on a two-dimensional electron
gas to confine the movement of electrons along one direction (Fig.
1a) \cite{zhang_guided_2009,jompol_probing_2009,hartmann_smooth_2010,wu_electronic_2011}.
Particularly, massless quasiparticles in graphene is an ideal platform
for the realization of such electron guide. The quasi-relativistic
linear energy dispersion in graphene allows the wavefunction of the
Dirac fermions travel with minimal distortion. Furthermore, it has been
demonstrated that high mobility \cite{dean_boron_2010} allows electrons
to be transmitted ballistically over several microns even at room
temperature \cite{mayorov_micrometer-scale_2011}. In addition,
graphene can be encapsulated between thin dielectric layers of hexagonal-boron
nitride (h-BN) \cite{wang_one-dimensional_2013}, providing tunable
electrostatic potential on the scale of a few nanometers, without
degradation of the mobility. Electrostatic gating has produced various
electron-optical elements, including lenses with negative refractive
index \cite{chen_electron_2016} and filter-collimator switches \cite{wang_graphene_2019}.

An ideal single mode electronic guide requires a deep potential well
with a width much smaller than the wavelength of electrons in order
to suppress scattering in the core of the waveguide \cite{allen_spatially_2016}.
The wavelength can reach around one hundred nanometers with experimentally
accessible densities, and it is therefore crucial to be able to place
extremely narrow gates close to the electron gas. The electronic modes
generated by such a 1-dimensional (1D) potential well are manifested
in the band structure of the graphene as branches similar to optical
modes, which are separated from the continuum up to the energy that
roughly corresponds to the depth of the potential well $U_{0}$ (Fig.
1b). Being isolated energetically, the guided modes are unlikely to mix with one
another. Moreover, they are predicted to propagate ballistically over exceptional distance \cite{shtanko_robustness_2018}. These modes form locally, at the center of the potential
well, such that they do not affect the overall graphene density of
states (DOS) but appear as resonances in the local density of states
(LDOS) close to the LDOS minimum which indicates the position of the
local Dirac point.

In this experiment, we use carbon nanotube (CNT) for creating 1D local
gate to create a guiding mode in graphene by generating a potential
well (Fig. 1a). The depth of the potential well can be continuously
adjusted by a voltage difference applied between the CNT and the graphene.
The width $d$ of the guided channel is roughly equal to the radius
of the CNT, around 1 nm, plus the thickness of h-BN separating the
CNT and graphene. The number of modes is then approximately given
by the ratio $U_{0}d/\hbar v_{F}$, where $v_{F}$ is the Fermi velocity,
and must, therefore, be of the order unity for a single mode waveguide
\cite{beenakker_quantum_2009,nguyen_quasi-bound_2009}. In principle,
this condition can be fulfilled for very wide and shallow potentials,
but for the mode to be well-defined it is necessary that the potential
depth is much greater than the fluctuations of chemical potential
caused by the disorder. This limitation explains in particular why
it is difficult to guide electrons in disordered graphene. For graphene
encapsulated in h-BN, these fluctuations are on the order of a few
meV \cite{xue_scanning_2011,yankowitz_van_2019}. In order to obtain
a single mode waveguide that is immune to disorder, the depth of the
potential well, therefore, needs to be around a few tens of meV, which
requires a width on the order of 10 nm (Fig. 1c). Such conditions
are very hard to fulfill with standard techniques of nanofabrication
but are conceivable using a gate made with a single-walled CNT in
close proximity to graphene. We also note that the linear dispersion
of Dirac fermion graphene is essential for our experiment. For non-relativistic
electrons in semiconductors, the criterion to have a single mode is
$U_{0}d^{2}\ll h^{2}/m$, where $m$ is the effective mass, leading
to a much shallower potential well ($\sim$meV) even for smaller width
$d<10$ nm (Fig.1c).

An optical image of one of our devices is shown in Fig. 1d. Graphene
is encapsulated between two layers of h-BN where the upper one is
only a few nm thick and on which a CNT is deposited. Since the CNT
diameter lies between 1 and 3 nm, and the thickness of our top h-BN
layer is only a few nanometers, the characteristic width of the well
is around 10 nm or less. We are thus able to drive the device into
a single guided mode formed in graphene beneath the CNT. The graphene
and CNT are both connected to their own electrodes which allows them
to be independently controlled and measured. The length of the waveguide
here is defined by the distance between electrodes connecting the
CNT, i.e. 500 nm. The details of fabrication are given in the supplementary
information \cite{SuppInfo}.

\begin{figure*}
\includegraphics[width=1\textwidth]{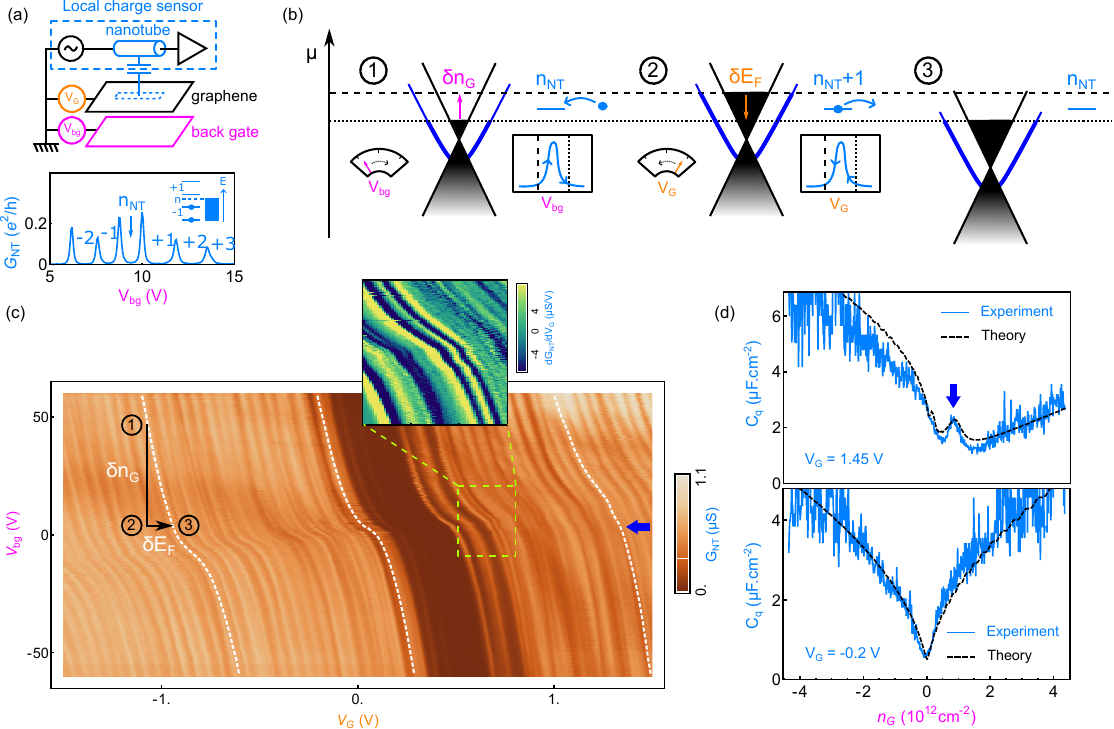}

\caption{\label{fig:Fig2}Graphene density of states measured with
the carbon nanotube (CNT). (a) Schematic of the measurement setup
(top) and CNT conductance measured as a function of the backgate showing
the Coulomb blockade behavior. All measurements presented
in this manuscript are performed at 1.6 K. (b) Operational
principle of the CNT sensor. (c) CNT conductance $G_{NT}$ versus
$V_{G}$ and $V_{bg}$. The wide dark brown area around
$V_{G}=0$ corresponds to the semiconducting gap of the CNT, in which
the latter is not charged. Inset shows $dG_{NT}/dV_{G}$ over a small
region in order to highlight a double kink corresponding to the Dirac
point followed by a guided mode resonance (blue arrow). (d)
Local quantum capacitance measured as a function of global charge
carrier density $n_{G}$ for two different voltage differences $V_{G}$
between the CNT and graphene.}
\end{figure*}

In addition to generating a potential well, the same CNT can also
be used as a local probe to measure the graphene LDOS utilizing the
capacitive coupling between CNT and the guided modes in the graphene.
Here we operate the CNT as a single electron transistor (SET), i.e.
a charge sensor \cite{martin_observation_2008}. Fig. 2a shows a schematic
of the measurement scheme where the electrostatic potential of CNT
SET can be controlled by both graphene gate voltage ($V_{G}$) and
the global back gate voltage ($V_{bg}$). When connected to metallic
electrodes and at sufficiently low temperature, a CNT generally enters
the Coulomb blockade regime and becomes sensitive to external charges
\cite{laird_quantum_2015}. By measuring the conductance $G_{NT}$ of the
CNT as a function of the gate voltage $V_{bg}$ or the potential applied
to the graphene sheet $V_G$, we observe a series of peaks corresponding
to the different electronic energy levels of the CNT (Fig. 2a, bottom
panel) each of which can contain one electron. These energy levels
can individually be used as local probes sensitive to the electrostatic
environment and therefore to the local charge density of graphene
located below the CNT. The operational principle of these probes,
inspired by direct measurements of Fermi energy performed in graphene
and bilayer graphene \cite{kim_direct_2012,lee_chemical_2014}, is
illustrated in Fig. 2b. When increasing the back gate potential $V_{bg}$,
we fill the graphene band structure by increasing the number of carrier
by $\delta n_{G}$ with the corresponding change of Fermi energy $\delta E_{F}$.
If the total electrochemical potential of graphene (electrostatic
potential added to the Fermi energy $E_{F}$) exceeds the energy of
one of the electronic states of the CNT, then the latter is also filled.
Subsequently, we lower the graphene electrostatic potential with $V_{G}$
and therefore reduce the energy of all the electrons in graphene by
an amount $\delta E_{F}$. If $\mu$, adjusted by a change of the graphene
bias $\delta V_{G}$, becomes lower than the energy of the same CNT
electronic level, it consequently empties and goes back to its original
state. By measuring the charge state of the CNT between each step,
it is then possible to deduce the energy change $\delta E_{F}=e\delta V_{G}$
corresponding a charge variation $\delta n_{G}$, where $e$ the charge
of an electron. This procedure yields the local quantum capacitance
of graphene 
\[
C_{q}=\frac{1}{e^{2}}\frac{\delta n_{G}}{\delta E_{F}}
\]
Note that the quantum capacitance at finite temperature is related to the compressibility
of a mesoscopic system $C_{q}=e^{2}\partial n_{G}/\partial\mu$, which can be
associated with the many body DOS \cite{yu_interaction_2013}. Since capacitive coupling between
graphene and CNT is strongly localized vicinity of the CNT, the measured
$C_{q}$ is proportional to the LDOS of graphene underneath of the
CNT. This quantity cannot be obtained with a global transport measurement. A remarkable aspect of this technique is that it provides an
absolute measurement of quantum capacitance without any scaling parameters
or adjustment of the origin of energies.

Fig. 2c shows the conductance of the CNT $G_{NT}$ as a function of $V_{bg}$
and $V_{G}$. For this particular device, the h-BN spacer between
CNT and graphene is only 4 nm thick, the measured peaks in the $G_{NT}$
exhibit trajectories in the $V_{bg}$ - $V_{G}$ plane that yield
the evolution of the Fermi energy as we described above. The slope
of these trajectories gives us directly the local quantum capacitance
$C_{q}$. When $V_{G}\approx0$, the potential difference between
the CNT and the graphene is small and, consequently, the potential
well generated by the presence of the CNT is shallow. The LDOS measured
(Fig. 2d) is then the one of bare graphene with a minimum at zero
energy, following $\left|n_{G}\right|^{1/2}$ on the hole and electron
sides. Note that $n_{G}$ denotes the global charge density of graphene
since $V_{bg}$ controls the charge density over the entire graphene
sheet. With the minimum of LDOS being very close to $n_{G}=0$, we
deduce that the doping underneath the CNT is low, suggesting a locally
low impurity levels. The minimum value of $C_q\sim0.5$ $\mu$F.cm$^{-2}$ also gives an estimation of the attainable minimal charge carrier concentration due to the charge puddle disorders $\sim10^{10}$ cm$^{-2}$.

\begin{figure*}
\includegraphics[width=1\textwidth]{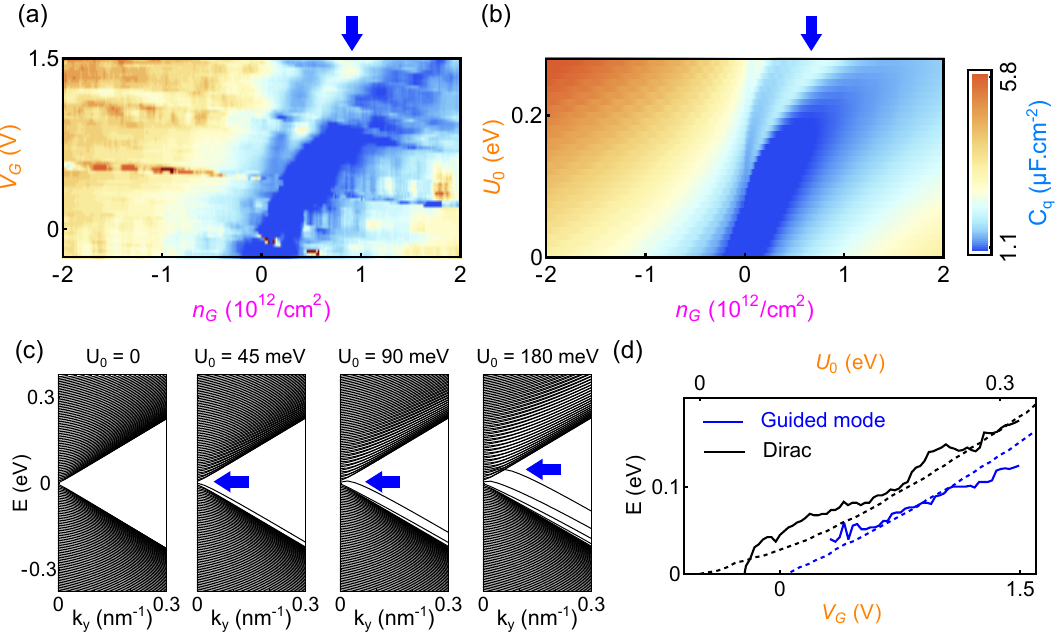}

\caption{\label{fig:Fig3}Potential depth dependence. (a) Evolution
of the graphene local density of states with the potential well depth
$U_{0}$ controlled by $V_{G}$. (b) Comparison with theoretical
calculations obtained from a tight-binding modeling. (c) Theoretical
evolution of the band structure for increasing $U_{0}$. (d)
Positions of local Dirac point and first guided mode as a function
of $V_{G}$ (experiment: continuous lines) and $U_0$ (theory: dashed lines). Here, the scales of $U_0$ and $V_G$ are adjusted to show the apparent linear relation between $U_0$ and $V_G$ in the accessible gate voltage range where the guided mode forms.}
\end{figure*}

As we generate a deeper potential well by increasing $V_{G}$, the
LDOS develops a more pronounced characteristic resonance, corresponding
to a single guided mode. Compared to the measurement performed at
$V_{G}=0$, the minimum of quantum capacitance has shifted from the
global charge neutrality point and towards the electron side ($n_{G}>0$),
as expected for a positive voltage applied on graphene while the CNT
is maintained at ground potential. The resonance lies between this
minimum and the global charge neutrality point of graphene ($n_{G}=0$),
a region where the doping caused by the potential well actually leads
to a NPN junction configuration. The appearance of this resonance
can be understood in the following manner: as a guided mode detaches
from the Dirac cone, it generates a peak in the LDOS due to the 1D
van Hove singularity appearing at the extrema of the single mode energy
dispersion $E(k_{y})$ where $k_{y}$ is the wave vector along the
CNT (see Fig. 3c). Our measurements are in excellent agreement with
numerical tight-binding simulations \cite{tworzydlo_finite_2008,hernandez_finite-difference_2012}
where the only fitting parameters are the depth and width of the potential
well (see supplementary information \cite{SuppInfo}). Theory predicts the appearance
of multiple successive modes that could give rise to additional resonances
\cite{jiang_tuning_2017}. However, due to presumably disorder induced
broadening, unambiguously identifying multiple resonances is challenging
within our experimental noise limit.

We observe a continuous evolution from bare graphene to a single mode
waveguide as we tune the potential depth $U_{0}$. Measurements of
Fig. 3a shows that the graphene LDOS appears to be dramatically affected
by tuning $U_{0}$, by changing the potential difference $V_{G}$
between the CNT and graphene becomes non-zero. At low $U_{0}$, it
is already clear that the minimum corresponding to the Dirac point
is less pronounced and that an asymmetry is formed between the electron
and hole sides. This evolution, also predicted by numerical simulations
(Fig. 3b), is due to the formation of closely packed guided modes
whose branches are too close to the continuum, preventing the development
of sharp resonances in the LDOS. Fig. 3c shows computed dispersion
relation as a function of $k_{y}$ momentum along the CNT direction.
A branch corresponding to the 1D guided mode gradually and continuously
separates from the Dirac cone as $U_{0}$ increases \cite{beenakker_quantum_2009}.
For larger $V_{G}$, we start to observe a resonance gradually increasing
in amplitude and shifting from the charge neutrality point. This reflects
the formation of a branch in the dispersion relation of graphene,
which becomes increasingly more detached from the continuum. The curvature
of this branch at its beginning becomes flat \cite{beenakker_quantum_2009}
until it acquires a minimum located around $k_{y}\approx1/d$, giving
rise to a sharp resonance in LDOS. In the relativistic Dirac fermionic
system, the 1D guide mode is expected to exhibit a potential strength
threshold for the appearance of the first guided mode \cite{nguyen_quasi-bound_2009,hartmann_quasi-exact_2014}.
While Fig. 3d suggests that indeed the appearance of a guided mode
starts at finite $U_{0}$, further experimental study with higher
resolution requires to prove such threshold behavior unambiguously.
Among all our devices, we were able to observe a single guided mode
in the ones with upper h-BN that are 6 nm or thinner. In devices made
with a thicker upper h-BN layer, from 10 to 100 nm, we were only able
to observe the asymmetry between electron and hole sides but no resonance
in the LDOS (see supplementary information). This experimental observation
confirms that we cannot create a robust single mode electronic waveguide
if the well is too wide and underlines the importance of the CNT for
the realization of the single guided mode.

Technological applications for guided modes are possible if the energy
separations between their branches and the continuum are sufficiently
large. Indeed, to make the information transmission robust along the
guide it is necessary to avoid processes that scatter electrons, leading
to loss of information. For applications operating at room temperature,
this energy must be well above thermal energy 25 meV. This separation
is directly given by the energy position of the resonance with respect
to the global Dirac point of graphene. Though our
measurement technique does not work at room temperature since it relies
on Coulomb blockade, we can see on the curve of Fig. 3d that we can
control this energy continuously up to approximately 0.1 eV, well
above thermal fluctuations at room temperature. This suggests that
such guided modes could have great potentials to be used as novel
electronic devices analogous to optical ones but where carriers of
information are electrons rather than light. Additional
measurements need to be performed at room temperature to confirm this
hypothesis. Complementary measurements based on infrared nano-imaging \cite{jiang_tunable_2016}, scanning tunneling spectrocopy \cite{jung_evolution_2011,chae_renormalization_2012,zhao_creating_2015}, or planar tunneling spectroscopy \cite{jung_direct_2017} with a 1D local gate created by a nanotube underneath the junction could also bring precious insight on the coherence, robustness and physical properties of these guided modes.
More generally, guided modes in Dirac materials are also of interest for plasmonics
applications \cite{jiang_tunable_2016,basov_polaritons_2016,ni_fundamental_2018}, ultrafast electronic \cite{huang_new_2018}, spintronics \cite{khosravi_dirac_2019}
or to be used as test-beds for relativistic simulation \cite{shytov_atomic_2007,wang_observing_2013,lu_frustrated_2019}.

\begin{acknowledgments}
We thank L. Levitov and J. Rodriguez-Nieva for discussion and M. O. Goerbig
for his helpful reading of our manuscript. The major part of this
work was supported by the Office of Naval Research (ONR N00014-16-1-2921).
P.K acknowledges a partial support from the Department of Energy (DOE
DE-SC0012260) for measurements. J.-D.P. acknowledges financial support
from Ecole Polytechnique. K.W. and T.T. acknowledge support from the
Elemental Strategy Initiative conducted by the MEXT, Japan and the
CREST (JPMJCR15F3), JST. 
\end{acknowledgments}

\clearpage
\widetext
\begin{center}

\textbf{\large Supplementary Information for: Guiding Dirac fermions in graphene with a carbon nanotube}

\vspace{1em}
{\large Austin Cheng , Takashi Taniguchi , Kenji Watanabe , Philip Kim, Jean-Damien Pillet}
\vfill
\end{center}

\clearpage

\setcounter{equation}{0}
\setcounter{figure}{0}
\setcounter{table}{0}
\setcounter{page}{1}
\makeatletter
\renewcommand{\theequation}{S\arabic{equation}}
\renewcommand{\thefigure}{S\arabic{figure}}

\section{Fabrication}

\subsection{Preparation of the circuit}

\begin{figure}
\begin{centering}
\includegraphics[width=0.6\textwidth]{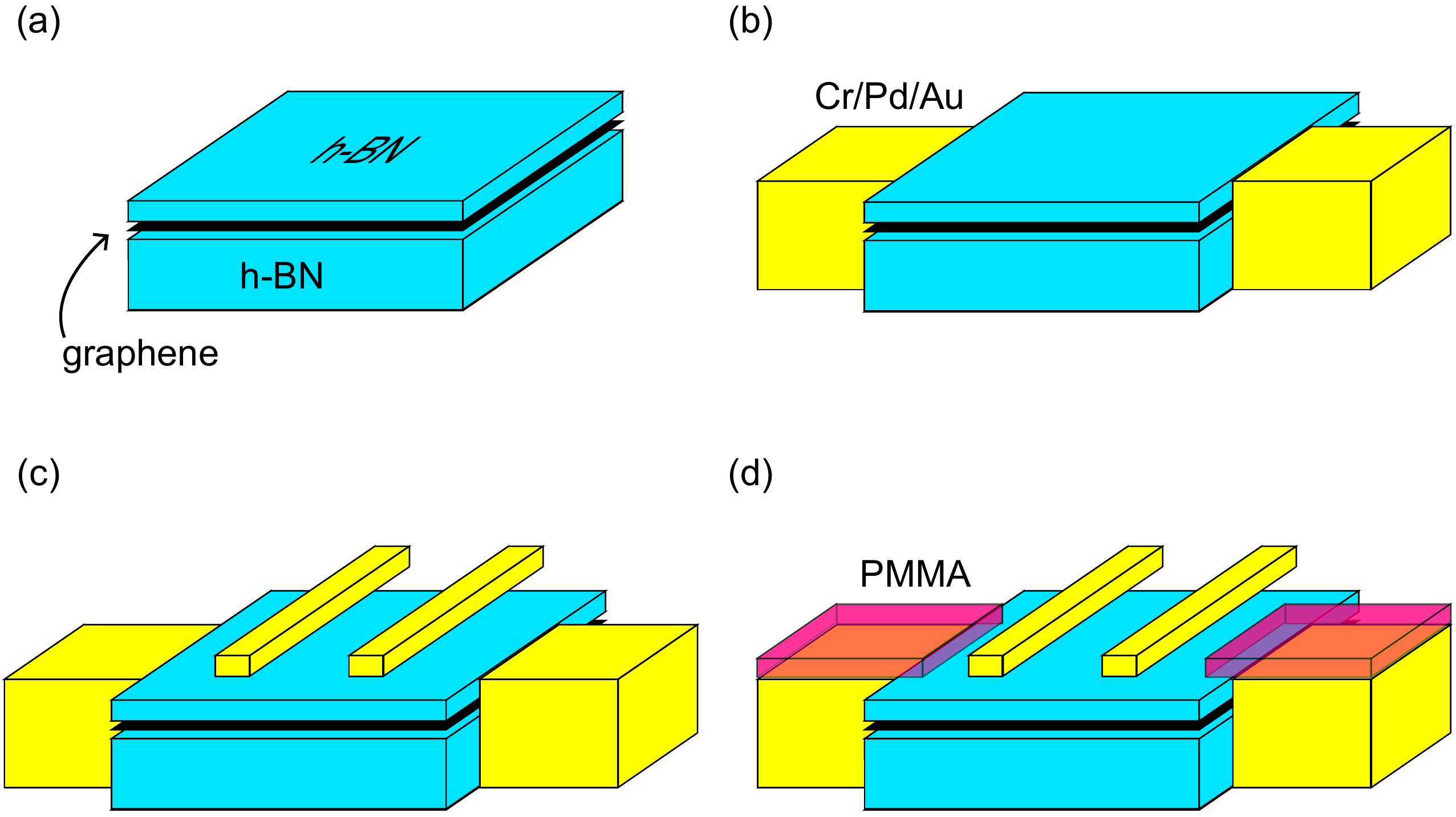}
\par\end{centering}
\caption{\label{fig:Preparation-of-stack}Sequence of preparation of the h-BN
encapsulated graphene for nanotube transfer. (a) h-BN/graphene/h-BN
sandwich. (b) The graphene is electrically connected after a first
step of lithography using RIE in order to expose the edges of the
graphene flake followed by a metallic evaporation. (c) Electrodes,
electrically isolated from the graphene by the top layer of h-BN,
are prepared for connection of the carbon nanotube. (d) The sample
is partially covered with resist, which helps the subsequent incorporation
of a carbon nanotube in the structure.}

\end{figure}

The sample shown in Fig. 1 of the main text is based on the initial
preparation of an h-BN encapsulated graphene \cite{wang_one-dimensional_2013}
(Fig. \ref{fig:Preparation-of-stack}a) on an n-doped silicon wafer
with 285 nm SiO2. The thickness of the top h-BN layer is chosen between
4 and 100 nm and the bottom one around 20 nm. We use standard technique
of e-beam lithography to design the electrodes contacting the graphene
flake. We first expose the edges of the graphene flake by reactive
ion etching through a resist mask and subsequently evaporate a metallic
trilayer Cr(5nm)/Pd(15nm)/Au(5nm) through the same mask (Fig. \ref{fig:Preparation-of-stack}b). A second step of lithography is then performed to design electrodes
(same metallic trilayer) on top of the top h-BN layer. These electrodes
are used to contact the carbon nanotube during the transfer step described
at the end of this section (Fig. \ref{fig:Preparation-of-stack}c).
The sample is covered with a 100nm thick layer of resist (PMMA A4
495K) except for areas of interest where we want the nanotube to connect
to electrodes. The resist helps on increasing the efficiency of the
transfer of the carbon nanotube (Fig. \ref{fig:Preparation-of-stack}d).

\subsection{Growth of nanotubes and characterization}

\begin{figure}
\begin{centering}
\includegraphics[width=0.4\textwidth]{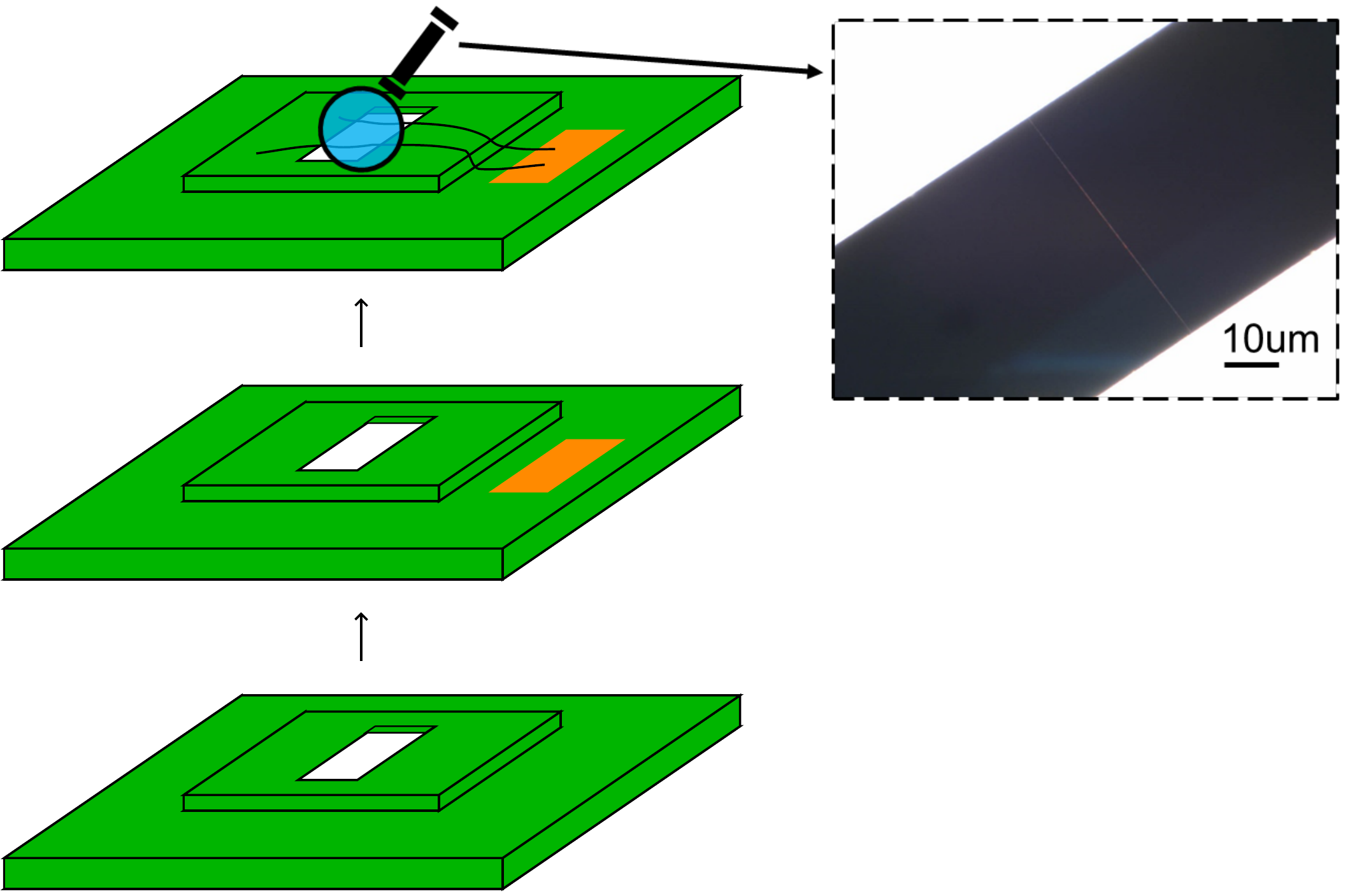}
\par\end{centering}
\caption{\label{fig:Growth}Sequence illustrating the growth of suspended carbon
nanotubes.}

\end{figure}

Carbon nanotubes are grown and characterized following the techniques
described in Ref. \cite{sfeir_probing_2004}. They are grown on $5\times5$mm$^{2}$
silicon chip with a slit in the center (see bottom of Fig. \ref{fig:Growth})
using standard technique of chemical vapor deposition. A catalyst
is deposited on one side of the slit (middle) such that carbon nanotubes
grow suspended (top). One of these nanotubes, suspended over a slit
that is 65$\mu$m wide and 1cm long, is shown in the optical picture
of Fig. \ref{fig:Growth}. It is covered with 30 nm of Au, so it can
be seen optically.

\begin{figure}
\begin{centering}
\includegraphics[width=0.6\textwidth]{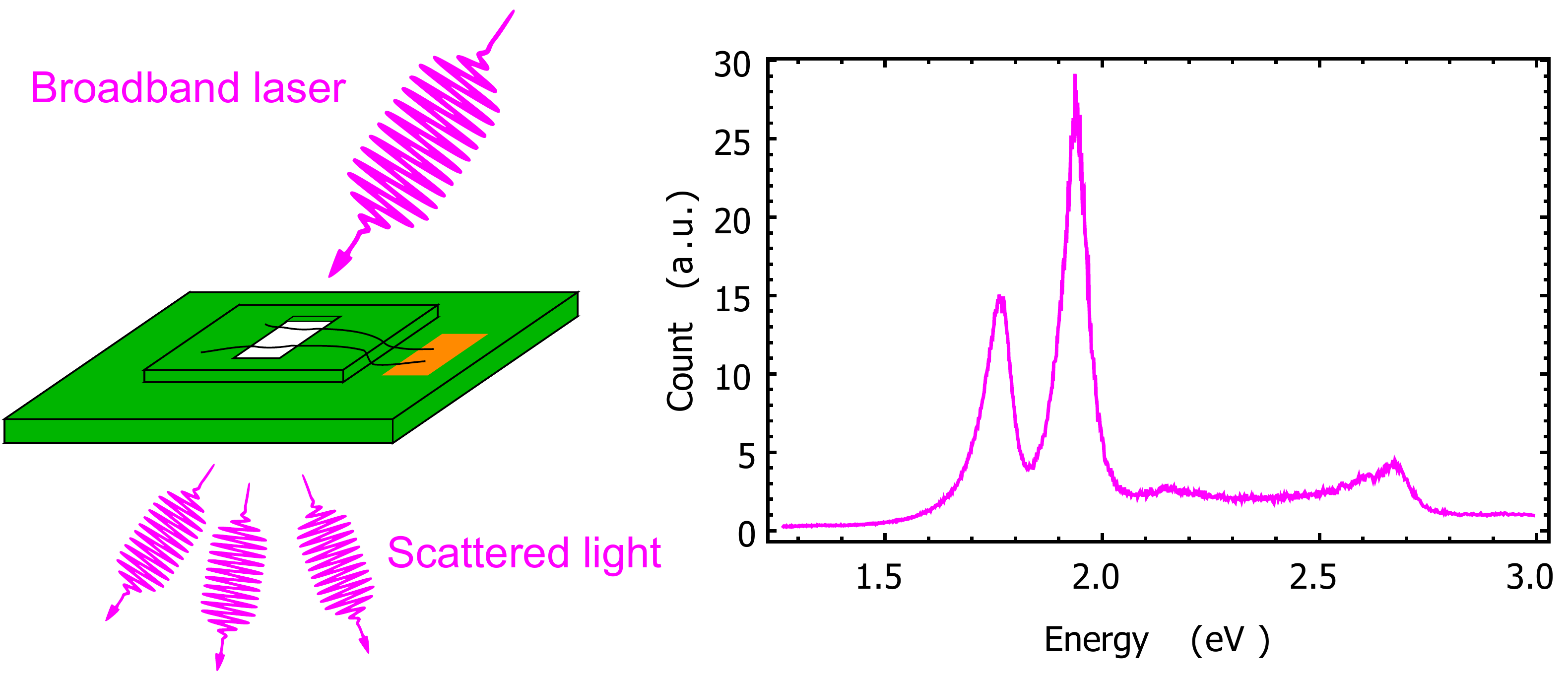}
\par\end{centering}
\caption{\label{fig:Rayleigh-scattering}Rayleigh scattering. Carbon nanotubes
are characterized when suspended over the slit. A broadband laser
is sent through the light and the scattered light is collected with
a detector (left). A typical spectrum is shown on the right, it gives
the nanotube chirality. In this example the nanotube was metallic
with a $\left(16,4\right)$ chirality.}

\end{figure}

After growth, carbon nanotubes can be characterized using Rayleigh
scattering. This identifies whether nanotubes are metallic or semiconducting
as illustrated in Fig. \ref{fig:Rayleigh-scattering}. Moreover it
also allows to measure the position of the carbon nanotube along the
slit such that it can be aligned with the circuit for subsequent transfer.

\subsection{Transfer}

\begin{figure}
\begin{centering}
\includegraphics[width=0.8\textwidth]{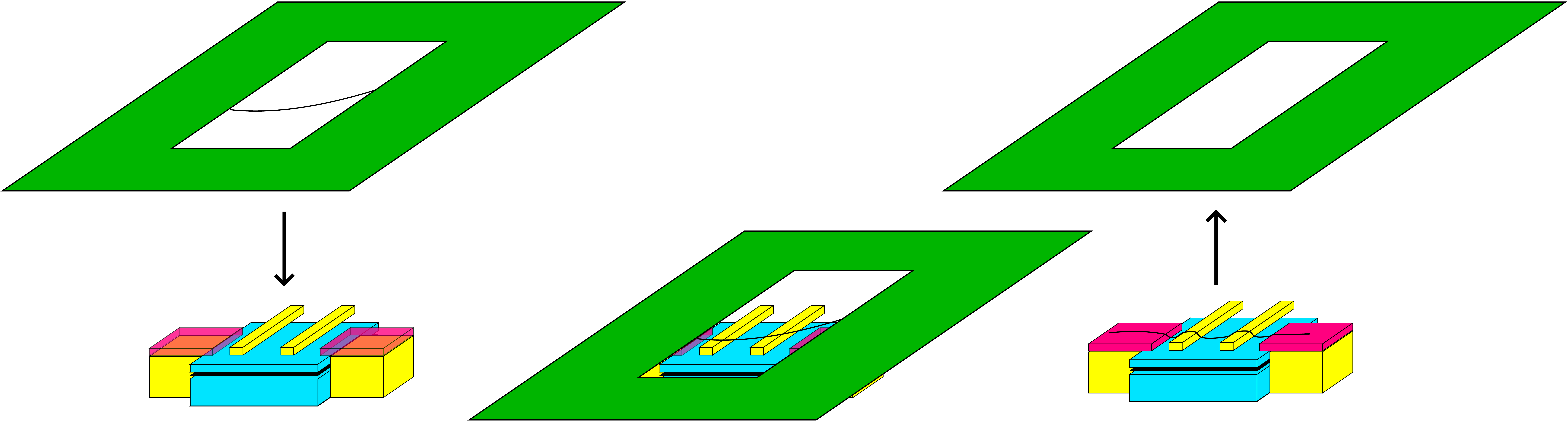}
\par\end{centering}
\caption{\label{fig:Transfer}Sequence illustrating the nanotube transfer.
The slit on which the nanotube is suspended is aligned with the h-BN
encapsulated graphene (left). It is then pressed onto the target chip
that is warmed up to 180$^{\circ}$C in order to melt the resist and
favor the transfer of the nanotube from one chip to another (center).
The two chips are then separated from each other and the nanotubes
is left onto its electrodes and over the h-BN/graphene/h-BN sandwich.}

\end{figure}

The incorporation of the carbon nanotube into the circuit is performed
by mechanical transfer \cite{huang_controlled_2005} similarly to
what is done to make h-BN encapsulated graphene. The slit is placed
above the circuit in order to align the nanotube with the area of
interest where we have designed dedicated electrodes. The slit is
pressed on the sample as shown in Fig. \ref{fig:Transfer}. When we
have good mechanical contact, we warm the chips up to 180$^{\circ}$C
for 5 minutes in order to melt the resist that helps the nanotube
transfer from the slit to the target chip. The two chips are then
slowly separated after they have cooled down to room temperature.

\section{Quantum capacitance from measurements}

\begin{figure}
\begin{centering}
\includegraphics[width=0.5\textwidth]{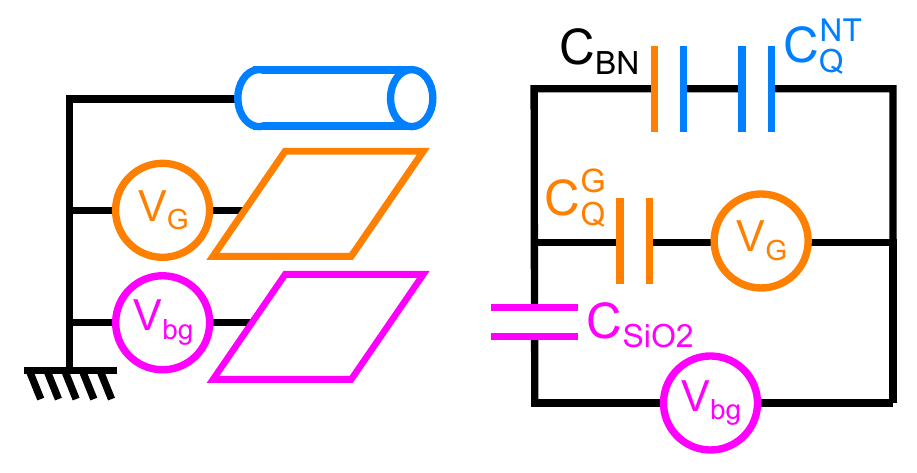}
\par\end{centering}
\caption{\label{fig:equivalent-circuit}Equivalent circuit of our hybrid carbon
nanotube-graphene devices}
\end{figure}

In Fig. \ref{fig:equivalent-circuit}, we describe our hybrid nanotube-graphene
device as a network of capacitances including geometric and quantum
capacitances in a fashion similar to what was done in Ref. \cite{kim_direct_2012}.
This schematic is equivalent to the following set of equations
\begin{equation}
\begin{cases}
V_{bg}-V_{G} & =e\frac{n_{G}+n_{NT}}{C_{SiO_{2}}}+\frac{E_{F}^{G}}{e}\\
-V_{G} & =-e\frac{n_{NT}}{C_{BN}}+\frac{E_{F}^{NT}-E_{F}^{G}}{e}
\end{cases}\label{eq:electrostatic-description}
\end{equation}
where, as defined in the text $V_{bg}$ is the voltage applied on
the back gate, $V_{G}$ is the voltage applied on the graphene flake,
$n_{G\left(NT\right)}$ is the number of carriers in the graphene
flake (resp. nanotube), $C_{SiO_{2}\left(BN\right)}$ is the capacitance
between the graphene flake and back gate (resp. nanotube) and $E_{F}^{G\left(NT\right)}$
is the Fermi energy of graphene. These equations can be obtained from
an electrostatic description of the circuit where the total energy
of the circuit $E_{tot}$ is given by
\[
E_{tot}=e^{2}\frac{\left(n_{G}+n_{NT}\right)^{2}}{2C_{SiO_{2}}}+e^{2}\frac{n_{NT}^{2}}{2C_{BN}}+\int_{-\infty}^{E_{F}^{G}}D_{G}\left(E\right)EdE+\int_{-\infty}^{E_{F}^{NT}}D_{NT}\left(E\right)EdE-eV_{bg}n_{bg}-eV_{G}n_{G}
\]
where $n_{bg}$ is the extra number of charge accumulated on the
back gate, $D_{G\left(NT\right)}\left(E\right)$ is the density of
states of graphene (resp. nanotube) as a function of energy $E$.
The total energy $E_{tot}$ contains five terms. The first two are
the electrostatic energies of the two geometric capacitors formed,
for the first one, by the back gate and the ensemble graphene-nanotube
and, for the second one, by the graphene flake and the nanotube. The
next two terms are the energies due to the fillings of electronic
levels in the nanotube and in the band structure of graphene. The
last two terms are the energy provided by the two voltage sources
applying respectively a potential $V_{G}$ and $V_{bg}$ on the graphene
and back gate. This energy is minimum when $\partial E_{tot}/\partial n_{NT}=0$
and $\partial E_{tot}/\partial n_{G}=0$ which, combined with the
condition that $n_{NT}+n_{G}=-n_{bg}$ since the circuit is a closed
system, lead to the system of equations \ref{eq:electrostatic-description}.
Note that here we have used the fact that $D_{G\left(NT\right)}\left(E_{F}^{G\left(NT\right)}\right)=\partial n_{G\left(NT\right)}/\partial E_{F}^{G\left(NT\right)}$.

In practice, Eq. \ref{eq:electrostatic-description} simplify using
the approximations $V_{G}\ll V_{bg}$, $n_{NT}\ll n_{G}$ and $E_{F}^{G}/e\ll V_{bg}$
that are essentially always valid since we apply tens of volts on
the back gate while we apply only hundreds of mV on the graphene flake,
the nanotube contains at best tens of electrons while the graphene
flake contains tens of thousands of electrons, and the Fermi energy
of graphene never exceeds a few hundreds of meV. By introducing the
nanotube quantum capacitance $C_{Q}^{NT}$, we end up with
\begin{equation}
\begin{cases}
V_{bg} & \approx e\frac{n_{G}}{C_{SiO_{2}}}\\
-V_{G} & \approx-e\frac{n_{NT}}{C_{NT}}+\frac{E_{F}^{G}}{e}
\end{cases}\label{eq:electrostatic-equations}
\end{equation}
with $C_{NT}^{-1}=C_{BN}^{-1}+\left(C_{Q}^{NT}\right)^{-1}$. From
these equations we see that, at a fixed number of charge in nanotube
($n_{NT}$ constant), we can deduce the number of charge in graphene
$n_{G}$ and its Fermi energy $E_{F}^{G}$ for given values of $V_{bg}$
and $V_{G}$.

As a consequence, we can write that along a trajectory made by an
electronic level of the nanotube (i.e. when the charge is fixed at
half an integer) in the $\left\{ V_{G},V_{bg}\right\} $ plane, we
have
\[
\frac{\partial V_{bg}}{\partial V_{G}}\approx\frac{e^{2}}{C_{SiO_{2}}}\frac{\partial n_{G}}{\partial E_{F}^{G}}\approx\frac{C_{q}}{C_{SiO_{2}}}.
\]
Since we have mesured independantly $C_{SiO_{2}}\approx12nF/cm^{2}$,
we can deduce directly the quantum capacitance of graphene from the
slope of these trajectories.

\section{Density of states calculation using a discretization of the massless
Dirac equation}

\subsection{Dirac Hamiltonian}

In order to describe our hybrid carbon nanotube-graphene devices,
we use the two-dimensional massless Dirac Hamiltonian given by 
\[
\hat{H}=v_{F}\left(\sigma_{x}\hat{p}_{x}+\sigma_{y}\hat{p}_{y}\right)+U\left(\hat{x}\right)
\]
where $v_{F}$ is the Fermi velocity, $\sigma_{x}$ and $\sigma_{y}$
the Pauli matrices. Due to the presence of the charged nanotube, the
electrostatic potential landscape $U\left(x\right)$ takes the shape
of a potential well that is invariant along the axis of the nanotube
(y-axis). For simplicity we choose a lorentzian potential though its
precise shape actually depends on how the electrons of graphene screen
the electric field generated by the nanotube. We write it as
\[
U\left(x\right)=\frac{U_{0}}{1+x^{2}/d^{2}}
\]
where $x=0$ corresponds to the position of the nanotube along the
x-axis, $U_{0}$ is the strength of the potential and $d$ is its
width. The latter depends on the radius of the nanotube as well as
the distance between nanotube and graphene given by the thickness
of the h-BN between them. We have also tested a logarithmic potential
$\propto-\log\left(\sqrt{x^{2}/d^{2}+1}\right)$, which corresponds
to the potential generated by a one-dimensional wire in a parallel
plane, but we have noticed no qualitative differences in the resulting
density of states.

\subsection{Discretized Dirac equations}

In order to calculate the density of states in the graphene flake,
we need to compute the eigenstates of this Hamiltonian $\Psi\left(x,y\right)$,
which obeys the massless Dirac equation
\[
\hat{H}\Psi\left(x,y\right)=E\Psi\left(x,y\right)
\]
where $E$ are the eigenenergies. Since our problem is invariant along
the $y$ direction, the operator $\hat{p}_{y}$ can be replaced by
a classical variable $p_{y}=\hbar k_{y}$ where $k_{y}$ is the projection
of the wavevector along this axis. As a consequence, the solutions
can be written as two-components spinors $\Psi\left(x,y\right)=e^{ik_{y}y}\left(\psi\left(x\right),\tilde{\psi}\left(x\right)\right)^{t}$,
which are plane waves along the y-axis.

We calculate $\psi$ and $\tilde{\psi}$ numerically using a discretization
of the Hamiltonian over a lattice whose points are separated by a
step $\Delta$. It is well known that a naive replacement of the derivative
by its discret equivalent might not preserve the hermiticity of the
Hamiltonian and cause a fermion doubling problem. To circumvent this
problem, we use a scheme called Susskind discretization \cite{tworzydlo_finite_2008,hernandez_finite-difference_2012}
\[
\partial_{x}\psi\rightarrow\frac{\tilde{\psi}_{m+1}-\tilde{\psi}_{m}}{\Delta}
\]
and
\[
\partial_{x}\tilde{\psi}\rightarrow\frac{\psi_{m}-\psi_{m-1}}{\Delta}
\]
where $\psi_{m}=\psi\left(m\Delta\right)$ and $\tilde{\psi}_{m}=\tilde{\psi}\left(m\Delta-\Delta/2\right)$
and $m$ is a relative integer such that $m\in\llbracket-N+1,+N\rrbracket$
for a flake of width $W=2N\Delta$. This means that we evaluate $\psi$
over the points of the lattice but $\tilde{\psi}$ at the midpoints.
The discrete version of the Dirac equation is then written
\begin{equation}
E\psi_{m}=-\frac{i\hbar v_{F}}{\Delta}\left[\left(\tilde{\psi}_{m+1}-\tilde{\psi}_{m}\right)+k_{y}\Delta\tilde{\psi}_{m}\right]+\frac{U_{0}}{1+\left[m\Delta/d\right]^{2}}\psi_{m}\label{eq:Dirac equation 1}
\end{equation}
and
\begin{equation}
E\tilde{\psi}_{m}=-\frac{i\hbar v_{F}}{\Delta}\left[\left(\psi_{m}-\psi_{m-1}\right)-k_{y}\Delta\psi_{m}\right]+\frac{U_{0}}{1+\left[\left(m-1/2\right)\Delta/d\right]^{2}}\tilde{\psi}_{m}\label{eq:Dirac equation 2}
\end{equation}
and we choose the boundary conditions $\psi_{-N}=\psi_{N+1}=\tilde{\psi}_{-N}=\tilde{\psi}_{N+1}=0$.
Such boundary conditions results in the formation of states on the
edge of the graphene flake, but it will not affect the local density
of states below the nanotube.

Solving Eq. \ref{eq:Dirac equation 1} and \ref{eq:Dirac equation 2},
we obtain branches of eigenenergies $E_{n}\left(k_{y}\right)$ with
corresponding eigenstates whose spinor components are $\psi_{m}^{E_{n}}\left(k_{y}\right)$
and $\tilde{\psi}_{m}^{E_{n}}\left(k_{y}\right)$. In practice, this
calculation is performed by first writing Eq. \ref{eq:Dirac equation 1}
and \ref{eq:Dirac equation 2} in the following form
\[
E\Psi_{E}\left(k_{y}\right)=\mathbb{H}\Psi_{E}\left(k_{y}\right)
\]
and then diagonalize numerically the matrix $\mathbb{H}$, which corresponds
to the Hamiltonian. Here, we have defined a $4N$-components vector
$\Psi_{E}=\left(\psi_{-N+1}^{E},\psi_{-N+2}^{E},...,\psi_{N}^{E},\tilde{\psi}_{-N+1}^{E},...,\tilde{\psi}_{N}^{E}\right)$,
and $\mathbb{H}=\mathbb{P}_{X}+\mathbb{P}_{Y}+\mathbb{U}$ with the
following $4N\times4N$ matrices
\[
\mathbb{P}_{X}=-\frac{i\hbar v_{F}}{\Delta}\left(\begin{array}{cc}
\mathbb{O} & \mathbb{D}_{+1}-\mathbb{I}\\
\mathbb{I}-\mathbb{D}_{-1} & \mathbb{O}
\end{array}\right),
\]
\[
\mathbb{P}_{Y}=-i\hbar v_{F}k_{y}\left(\begin{array}{cc}
\mathbb{O} & \mathbb{I}\\
-\mathbb{I} & \mathbb{O}
\end{array}\right)
\]
and
\[
\mathbb{U}=U_{0}\left[\left(\begin{array}{cc}
\mathbb{I} & \mathbb{O}\\
\mathbb{O} & \mathbb{I}
\end{array}\right)+\frac{\Delta^{2}}{d^{2}}\left(\begin{array}{cc}
\mathbb{X} & \mathbb{O}\\
\mathbb{O} & \mathbb{X-I}/2
\end{array}\right)^{2}\right]^{-1}.
\]
In these expressions we have introduced $2N\times2N$ matrices where
$\mathbb{O}$ is a matrix full of zeros, $\mathbb{I}$ is the identity
matrix, $\mathbb{D}_{\pm1}$ are matrices in which all the coefficients
are zero except on the first upper (resp. lower) diagonal where all
the coefficients are equal to 1, and $\mathbb{X}=diag\left(-N+1,...,N\right)$
is a diagonal matrix which refers to the the position along the x-axis.

Diagonalizing the matrix $\mathbb{H}$, we obtain the eigenvalues
$E_{n}\left(k_{y}\right)$ and the corresponding eigenstates $\Psi_{E_{n}}\left(k_{y}\right)$. Note that in this approach, we suppose that the doping remains moderately small which consequently limits the screening effect. We therefore neglect the non-linear response of graphene to the electrostatic potential generated by the nanotube \cite{jiang_electronic_2015}.

\subsection{Global and local density of states calculations}

The global density of states in graphene $DOS\left(E\right)$ is given
by 
\[
DOS\left(E\right)=\frac{1}{\pi}\sum_{k_{y}}\sum_{E_{n}}\frac{2\gamma}{\left[E-E_{n}\left(k_{y}\right)\right]^{2}+\gamma^{2}}
\]
where the factor $2$ accounts for the spin degree of freedom and
where we have introduced a phenomenological broadening $\gamma$ for
each electronic level of energy $E_{n}\left(k_{y}\right)$ in order
to smooth the density of states. In practice, we choose $\gamma=0.01\times\hbar v_{F}/\Delta$
in our calculations, which roughly corresponds to the distance between
two energy levels. The total density of states is obtained by summing
over all the eigenenrgies of $\mathbb{H}$ ($4N$ in total) for a
given $k_{y}$ and then by summing over $k_{y}$. Here we consider
a graphene flake width $W=2N\Delta$ and length $L$ such that we
consider $k_{y}$ to be quantized quantized in steps of $\pi/L$ in
the interval $\left[-\frac{\pi}{\Delta},\frac{\pi}{\Delta}\right]$.
In our calculations, we choose $\Delta=1$ nm, $N=200$ and $L=50\pi\Delta$.

The local density of states $LDOS\left(E\right)$ below the nanotube
is obtained in a similar fashion but taking into account the spatial
distribution of the wavefunctions
\[
LDOS\left(E\right)=\frac{1}{\pi}\sum_{k_{y}}\sum_{E_{n}}\frac{2\gamma}{\left[E-E_{n}\left(k_{y}\right)\right]^{2}+\gamma^{2}}\frac{1}{Tr\left(\mathbb{M}\right)\Delta L}\Psi_{E_{n}}\left(k_{y}\right)^{t}\mathbb{M}\Psi_{E_{n}}\left(k_{y}\right)
\]
where the matrix $\mathbb{M}$ accounts for the small region below
the nanotube over which the LDOS is measured. We choose this region
to have the same width $d$ as the electrostatic quantum well created
by the same nanotube such that $\mathbb{M}$ can be written as
\[
\mathbb{M}=\left[\left(\begin{array}{cc}
\mathbb{I} & \mathbb{O}\\
\mathbb{O} & \mathbb{I}
\end{array}\right)+\frac{\Delta^{2}}{d^{2}}\left(\begin{array}{cc}
\mathbb{X} & \mathbb{O}\\
\mathbb{O} & \mathbb{X-I}/2
\end{array}\right)^{2}\right]^{-1}.
\]
where we have assumed that the sensitivity of the nanotube decreases
with distance following a lorentzian decay. Note that $Tr\left(\mathbb{M}\right)\Delta L$
corresponds to the surface over which the nanotube measures the local
density of states, which means that $LDOS\left(E\right)$ is a local
density of states per unit area.

\subsection{Fitting parameters $U_{0}$ and $d$}

In our simulation, we need only two fitting parameters to describe
quantitatively the density of states we measure. The first one is
the depth of the potential $U_{0}$ that we experimentally control
with the voltage difference applied between the carbon nanotube and
graphene $V_{G}$. The second is the width $d$ of the potential which
is roughly given by the radius of the nanotube plus the thickness
of the h-BN spacer between nanotube and graphene. However, this only
gives us an approximate value as the shape of the potential well generated
by the nanotube will be affected by the screening of the graphene
electrons. We obtain good agreement between theory and experiments
using a width of 10 nm, that is to say $d=10\Delta$ in our simulations
shown in Fig. 3b of the main part of the manuscript.

\section{Devices with wider potential well}

\begin{figure}
\includegraphics[width=1\textwidth]{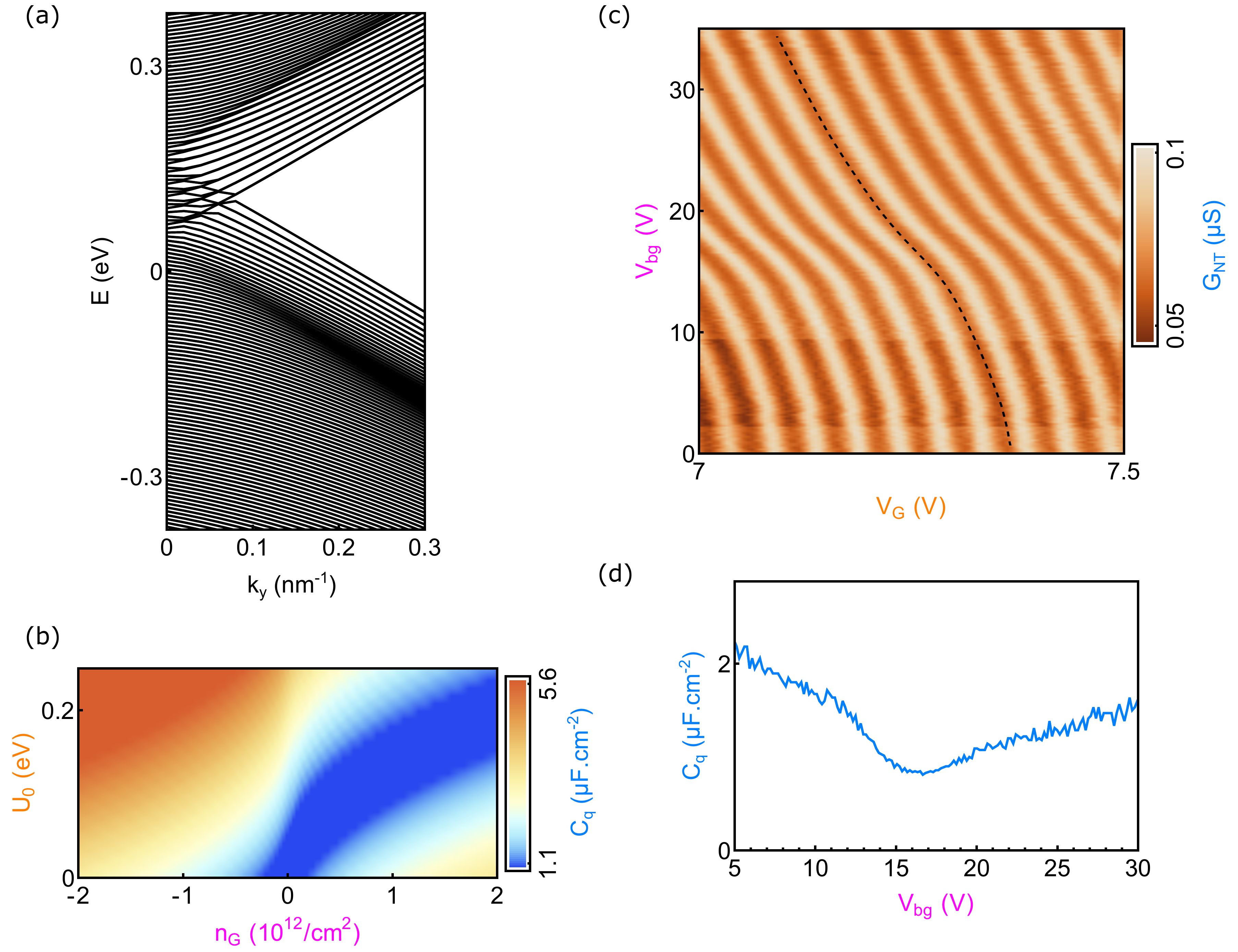}

\caption{\label{fig:Wider Potential}Wider potential well. (a) Band structure of graphene for a deep and wide potential well. (b) Simulation of quantum capacitance measurement for a device with a wide potential well. (c) Measurement of nanotube conductance for a device with a 30 nm thick h-BN separator between nanotube and graphene. (d) Measured quantum capacitance of graphene for a deep and wide potential well.}

\end{figure}

Fig. \ref{fig:Wider Potential} shows simulations and measurements
for devices with wider potential wells. The shown data come from a
device with 30 nm thick h-BN separating the CNT and Gr. We observe
qualitatively similar behaviors in devices with h-BN that is 10 nm
or thicker.

The simulations are performed for a potential that is 100 nm wide.
Fig. \ref{fig:Wider Potential}a shows the dispersion relation of
Gr underneath the CNT when the potential is at $U_{0}=0.45$ eV. We
observe that the branches are essentially indistinguishable from the
continuum. Unlike the single mode which is well isolated from the
continuum, these multi-modes can couple with one another as well as
with the continuum easily, which makes them poorly guided.

Figure \ref{fig:Wider Potential}b shows the numerically calculated
$C_{q}$. We do not see resonances developing due to the fact that
branches detaching from the continuum are too close from each other
and the continuum. Figure \ref{fig:Wider Potential}c shows a carbon
nanotube conductance measurement for the device around a high gate
potential ($V_{G}=7.25$ V). The trajectories of the conductance peaks
form smooth S curves which represent the Dirac point and there are
no \textquotedblleft kinks\textquotedblright{} seen elsewhere. This
signifies the absence of resonance in the density of states as shown
in \ref{fig:Wider Potential}d.

However, we do observe an asymmetry developing between the elecron
and hole sides and a smoothing of the Dirac point, meaning that the
electric field generated by the nanotube affects the graphene density
of states. The fact that we do not see resonances for wider potentials
is in agreement with theory and an indication of the formation of
several modes rather than a single guided mode. This provides further
evidence that significantly sharp potential wells are required for
the realization of a single mode electron guide.

\end{document}